\documentclass{article}
\usepackage[english]{babel}
\usepackage[utf8]{inputenc}
\usepackage[dvips]{graphicx}
\usepackage{authblk}
\usepackage{url}

\title{The thermodynamic cost of measurements}
\author{Léo Granger}
\author{Holger Kantz}
\affil{Max Planck Institut für Physik komplexer
Systeme, Nöthnitzerstr. 38, D-01187, Dresden, Germany}

\begin{document}

\maketitle
Email: \url{granger@pks.mpg.de}

\begin{abstract}
The measurement of thermal fluctuations  provides information
about the microscopic state of a thermodynamic system and
can be used in order to extract work from a single heat bath
in a suitable cyclic process.
We present a minimal framework for the modeling of a
measurement device and we propose a protocol for the
measurement of thermal fluctuations.
In this framework, the measurement of
thermal fluctuations naturally leads to the dissipation of
work. We illustrate this framework on a simple two states
system inspired by the Szilard's information engine.
\end{abstract}

\section{Introduction}

In his seminal paper of 1929, Leó Szilard was the first to
point out the role of information in statistical
thermodynamics \cite{szilard29}. 
Recently, experimental and theoretical work have specified
the relation
between information and dissipated work 
in the thermodynamics of small systems
\cite{toyabe2010, abreu2011,hasa1, hasa2,esp2011, 
sagawa2008, sagawa2010, sagawa2011}. 
In its traditional formulation,
the second law of thermodynamics states that the
average work $W$ needed to change the state of a system in contact
with a heat bath is bounded from below by the difference in
free energy of the final and the initial states:
\begin{equation}
W\geq \Delta F.
\end{equation}
In the presence of measurement and  feedback during the
process, the second law
has to be extended in order to include the information
obtained through the measurement and the bound on the work
to perform is lowered \cite{sagawa2008, sagawa2010, sagawa2011, horo2010,
pon2010}:
\begin{equation}
W\geq \Delta F - kTI,
\end{equation}
where $I$ is the mutual information of the system state and
the measurement outcome, $k$ is Boltzmann's constant and $T$ is the
temperature of the heat bath.

A striking consequence of
this relation is the theoretical possibility to extract work 
out of a single heat bath during a cyclic process. 
The second law of thermodynamics prohibits such processes.
In order to re-establish the second law it is therefore essential  
that the acquisition (and/or processing) of an amount $I$ of
information leads to the dissipation of at least $kTI$ of
work.

Landauer and Bennett indeed focused on the processing of the
information \cite{landauer, bennett}. They argue that, in
order to utilize some information, one has to record it on
some memory device and eventually to erase it. Landauer's
principle sates, that this erasure step is necessarily
accompanied by a minimum amount of entropy production
sufficient to balance the entropy reduction due to the
feed-back process. This principle has been the subject of a
lot of studies, see e.g. \cite{fahn, sagawa2009}.

The aim of this paper is to propose a framework for the
modeling of a measurement device. Under thermodynamically
consistent assumptions, the measurement of {\em thermal
fluctuations} naturally leads to dissipation in a way
similar to Landauer's principle.
The basic assumptions about the measurement device are the
following: it should be a thermodynamic system subject to
thermal fluctuations and it should receive information from
the system on which the measurement is performed. The first
assumption implies that the measurement errors should at
least include the thermal fluctuations inside the
measurement device. The second assumption implies that the
measurement device is {\em driven} by the original system.
In our framework, it is this driving that is responsible for
the entropy production inside the measurement device

\section{Modeling the measurement device}

We wish to measure a certain quantity $x$. Here, $x$ is
thought of as a random variable distributed according to some
probability distribution $p(x)$. 
The knowledge that we have about $x$ is given by the {\em
Shannon entropy} of $p(x)$ given by \cite{cover}:
\begin{equation}\label{eq:shannon}
S[p(x)] = -\sum p(x)\log p(x).
\end{equation}
The lower it is, the more information we have about $x$.
By measuring $x$, we mean
acquiring some information about a single realization of
this random variable. Let $y$ be the result of the
measurement, distributed according to the conditional
distribution $p(y|x)$ for fixed $x$. If we know the value
of $y$, then our knowledge about the value of $x$ changes
and $x$ is distributed according to the conditional
probability distribution:
\begin{equation}\label{eq:xknowingy}
p(x|y) = \frac{p(y|x)p(x)}{p(y)},
\end{equation}
where $p(y) = \sum_{x} p(y|x)p(x)$ is the marginal
distribution of $y$, i.e. the a priori probability to observe
outcome $y$. The entropy of $x$ after observing $y$ is the
Shannon entropy of $p(x|y)$:
\begin{equation}
S[p(x|y)] = - \sum_x p(x|y)\log p(x|y).
\end{equation}
This quantity depends on the measurement outcome $y$.
On average over $y$, it is smaller than the Shannon entropy
of $p(x)$ given by eq.~(\ref{eq:shannon}), meaning that
knowing the value of $y$ increases our information about
$x$. The average decrease of entropy of $x$ upon knowing the
value of $y$ is the {\em mutual information} between $x$ and
$y$ \cite{cover}:
\begin{eqnarray}\label{eq:info}
I & = & S[p(x)] - \sum_y p(y)S[p(x|y)]\nonumber \\ 
  & = & \sum_{x,y} p(x,y)\log\frac{p(x,y)}{p(x)p(y)},
\end{eqnarray}
where $p(x,y) = p(y|x)p(x) = p(x|y)p(y)$ is the joint
distribution of $x$ and $y$. The mutual information $I$ is
positive and it is zero if and only if $x$ and $y$ are
independent.

If $x$ is the microscopic (or mesoscopic) state of a
thermodynamical system in equilibrium with a heat bath at
temperature $T$, then the information obtained can be used
to extract heat from the heat bath and convert it into work
\cite{szilard29, toyabe2010, abreu2011, sagawa2008, 
sagawa2010, sagawa2011}. More precisely, let $x$ be the
microscopic state (or micro-state) of a thermodynamic
system $S$ in contact
with a heat bath at temperature $T$ and let $p(x)$ be its
equilibrium (canonical) distribution. If one knows the value
of $y$, then the micro-states of $S$ are distributed according
to $p(x|y)$ given by eq.~(\ref{eq:xknowingy}). This
distribution is a non-equilibrium one and exploiting the
relaxation of $S$ to equilibrium allows one to extract a maximum
average amount of work linked to the {\em Kullback-Leibler distance}
or {\em relative entropy} of the non-equilibrium
distribution $p(x|y)$ and the equilibrium one $p(x)$
\cite{hasa1,esp2011}:
\begin{equation}
W_{\rm max}(y) = kT \sum_x p(x|y)\log\frac{p(x|y)}{p(x)},
\end{equation}
where $k$ is Boltzmann's constant. On average over $y$, one
obtains:
\begin{equation}
W_{\rm max} = \sum_y p(y)W_{\rm max}(y) = kTI,
\end{equation}
where $I$ is the mutual information between $x$ and $y$
given by eq.~(\ref{eq:info}).

The measurement device should be a physical system obeying the
laws of thermodynamics. Moreover, it should receive
information from the original system $S$. These
considerations lead us to consider the measurement device as
a thermodynamic system $M$ and the measurement outcome $y$
as a micro-state of $M$.
The energy levels of $M$ depend on the
value $x$ to be measured in such a way, that $p(y|x)$ is
the equilibrium distribution of $y$. Denoting by $E_M(y|x)$ the
energy of $M$ when $S$ is in state $x$ and $M$ in state $y$,
we have:
\begin{equation}\label{eq:thermerr}
p(y|x) = \exp\left(-\frac{E_M(y|x) - F_M(x)}{kT}\right),
\end{equation}
where $F_M(x) = -kT\log\sum_y\exp(-E_M(y|x)/kT)$ is the
equilibrium free energy of $M$ given $x$. 
Every time the value of $x$ changes, $M$ is driven away
from equilibrium.
Furthermore, we suppose the relaxation time of $M$ to be
much smaller than the relaxation time of $S$,
 so that $M$ always
has the time to relax towards the canonical distribution
(\ref{eq:thermerr}) before the value of $x$ changes. 

During the measurement, $M$ is in a {\em probabilistic
mixture} of macro-states \cite{Ladyman2008}. By this, we
mean, that the macroscopic state of $M$ is random. This is
so because it depends on $x$ which is itself random. The
measurement errors come form the difficulty to distinguish
different macro-states of $M$ upon seeing one realization of
$y$. In fact, different $p(y|x)$ for different values of $x$
may overlap, meaning that different values of $x$ may be
compatible with one value of $y$.

\section{Measuring the state of a two levels
system}\label{sec:setup}

We will start with a simple example inspired by the
Szilard's engine \cite{szilard29,horo2011} in order to introduce the
measurement protocol, and then we will consider a more
general case. The aim is to measure the state of a two
levels system $S$ in contact with a heat bath at constant
temperature $T$, possibly with measurement errors.
The information obtained through the measurement is then
used in
order to extract some heat out of the heat bath and convert
it into work. We denote by $x\in\{1,2\}$ the state of $S$
and by $y\in\{1,2\}$ the result of the measurement.
Initially, $S$ is in equilibrium with the heat
bath and the energies of its levels are equal so that it has
equal probability to be in one state or the other. At some
point, we measure the state of $S$ with success probability
$p\geq 1/2$: 
\begin{equation}\label{eq:measerr}
p(y|x) = \left\{
\begin{array}{lcl}
p & {\rm if} &y=x\\
1-p & {\rm if} &y\neq x.
\end{array} \right.
\end{equation}
By a cyclic process depending on the measurement
outcome one can extract at most $W_{\rm max}=kT I$ of work,
where $I$ is the information gained through the measurement
 \cite{sagawa2010,sagawa2011, horo2011}:
\begin{equation}\label{eq:info2}
I(p) = p\log\frac{p}{1/2} + (1-p)\log\frac{1-p}{1/2}
\end{equation}
for this specific system.

The measurement device $M$ is a two level system as well
since the measurement has two possible outcomes. Its energy
levels are separated by a gap linked to $p$ by 
\begin{equation}\label{eq:gap}
\Delta E = -kT\log\frac{p}{1-p},
\end{equation}
so that $p(y|x)$ given by eq.~(\ref{eq:measerr}) is the
equilibrium distribution for $y$. In other words:
\begin{equation}\label{eq:em}
E_M(y|x) = 
\left\{
\begin{array}{lcl}
-kT\log p + F_M & {\rm if} & y=x\\
-kT\log(1-p) + F_M & {\rm if} & y\neq x,\\
\end{array}
\right.
\end{equation}
where $F_M$ is the free energy of $M$.
Hence, during the measurement, each macroscopic state of $M$
is labeled by a state of $S$.

The measurement process consists of three steps:
\begin{enumerate}
\item Initially, $M$ is independent of $S$ and $S$ occupies
one of its two states with equal probability.
\item\label{it:contact} At some point $M$ is ``put in contact" 
with $S$ and its
energies are switched to the values given by eq.~
(\ref{eq:em}).
It relaxes immediately towards equilibrium, so
that it is described by the conditional distribution
(\ref{eq:measerr}).
Having a look
at the value of $y$ yields on average information $I(p)$,
given by eq.~(\ref{eq:info2})
about the value of $x$.
\item\label{it:sep} Finally, $M$ is ``decoupled" from $S$ 
in the sense that
it does not anymore receive information from it. A protocol
fed back with the result of the measurement is performed
on $S$ yielding a maximum average amount of $kTI(p)$ of
work.
\end{enumerate}
The problem is now to find the quantity of work performed on
$M$ during this cycle of transformations.
But before doing so, let us make one short remark. It is
important that $M$ does not stay in ``contact'' with $S$
during the whole process. If it did, then each time $S$
would jump from one state to the other because of a thermal
fluctuation, work would be performed on $M$. Hence, in order
to dissipate the smallest possible amount of work, it is
important to make the contact as short as possible. It
should be just long enough to yield the desired information,
but not longer.

We first need to specify the macroscopic state of $M$ in
steps 1 and 3, that is, when it is not coupled to $S$.
For the process to be cyclic, we require
these states to be the same. Consider two possibilities:
\begin{itemize}
\item $M$ is in some standard state, with equal energies
$E_M^0$.
\item When $M$ is decoupled from $S$, it is just left as it
is. In other words, it is in a {\em statistical mixture} of
macroscopic states: with probability 1/2 it is in a state
with energies $E_M(y|1)$ and with probability 1/2{} in a
state with energies $E_M(y|2)$.
\end{itemize}

In the first case, work is performed on $M$ in step 
\ref{it:contact}
and \ref{it:sep}, 
i.e. when $M$ is put in contact with $S$ and when it
is separated from $S$. When $M$ is put in contact to $S$,
its energy levels are instantly moved from $E_M^0$ to
$E_M(y|x)$. 
The averaged work dissipated
when one instantly changes the energies of a system
is given by $kT$ times the  Kullback-Leibler
distance between the initial equilibrium distribution and
the final one. This is a special case of the
Kawai-Parrondo-Brock equality \cite{kawai2007, gomez2008}.
We will make use of this relation all along this article to
calculate the work performed at each steps of the process.
the work performed when $M$ is put in contact with $S$ is
thus given by:
\begin{equation}
W_{\rm cont} = \frac{kT}{2}\left( \log\frac{1/2}{p} +
\log\frac{1/2}{1-p}\right) +\Delta F_M,
\end{equation}
where $\Delta F_M=F_M-F_M^0$ is the difference in the
free energy of
$M$ before and after the contact. When $M$ is separated from
$S$, the work performed on average is:
\begin{equation}
W_{\rm sep} = kT\left( p\log\frac{p}{1/2}+
(1-p)\log\frac{1-p}{1/2}\right) -\Delta F_M.
\end{equation}
The first term on the right hand side of this equation is
the mutual information between $x$ and $y$, eq.
(\ref{eq:info2}). Thus
\begin{equation}
W_{\rm sep} = kTI(p) -\Delta F_M.
\end{equation}
The overall work performed is:
\begin{eqnarray}\label{eq:measwork}
W_{\rm tot} & = & W_{\rm cont} + W_{\rm sep}\\
 & = & kTI(p) +\frac{kT}{2}\left(\log\frac{1/2}{p} +
\log\frac{1/2}{1-p}\right).\nonumber
\end{eqnarray}
Obviously, the work performed on $M$ is greater than the
work that can be extracted from $S$ using the information
provided by the measurement because:
\begin{equation}
D(p) = \frac{1}{2}\left(\log\frac{1/2}{p} +
\log\frac{1/2}{1-p}\right) \geq 0
\end{equation}
with equality if and only if $p=1/2$, i.e. when the
measurement does not provide any information.

In the second case, work is only performed during the
contact since $M$ is left unchanged after the measurement.
During the contact, with probability 1/2 $M$ does not change
and no work is performed on it and with probability 1/2, the
energies of the levels of $M$ are exchanged. The average
work performed is:
\begin{equation}
W' = \frac{kT}{2}\left( p\log\frac{p}{1-p} +
(1-p)\log\frac{1-p}{p}\right).
\end{equation}
One can show that $W'=W_{\rm tot}$, meaning that it makes no
difference whether one uses a definite standard state or
not.
In the next section, we
will generalize this result to the general case of a
measurement with an arbitrary number of outcomes with an
arbitrary distribution.

\begin{figure}
\centering
\includegraphics{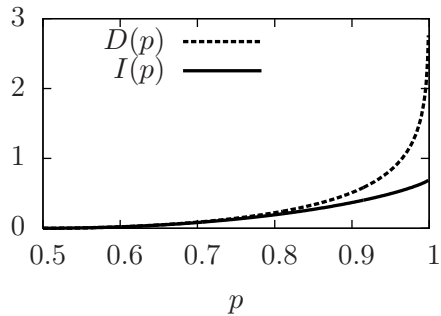}
\caption{The two different contributions $D(p)$ and $I(p)$
to the work performed on $M$ during the measurement process
 as a function of the probability $p$
of a successful measurement. $D(p)$ is the contribution due
to the contact and $I(p)$ is the contribution due to the
separation.}
\label{fig:plot}
\end{figure}

As can be seen on figure \ref{fig:plot}, $D(p)\geq I(p)$.
Hence, the contribution of the contact step to the
dissipated work is greater than the contribution of
separation step.

\section{Generalization}

The situation is similar in the general case. Consider the
setup depicted in section \ref{sec:setup}: $S$ is a
thermodynamic system in contact with a heat bath at
temperature $T$ and $p(x)$ its canonical distribution. The
measurement device is also a thermodynamic system in contact
with the heat bath and the energy of its micro-state either
have some standard value $E_M^0(y)$ or are given by eq.
(\ref{eq:measerr}), i.e. they are such that $p(y|x)$ is the
canonical distribution for $M$ for a given value of $x$.
Since the
states of $M$ correspond to the possible measurement
outcomes, it would make no sense that $y$ can take
more values than $x$.
However, formally, the following
derivation is still valid in that case.

As in the previous section, we consider two different
protocols for the measurement. Either the energy levels of
$M$ are driven from the standard values $E_M^0(y)$ to the
values corresponding to the measurement, $E_M(y|x)$ and then
back to the standard values. Or the energies of $M$
initially have the
values given by the previous measurement, i.e. $E_M(y|x')$ with
probability $p(x')$, $x'$ being the state of $S$ during the
previous cycle, and are driven to the values corresponding to
the actual measurement. We will see that for a suitable
choice for $E_M^0(y)$, the two protocols give the same value
for the work performed on $M$, as in the case of a two
states system.

We set the values of $E_M^0(y)$ so that the marginal
distribution of the measurement outcome $y$ is an
equilibrium one:
\begin{equation}
E_M^0(y) = -kT \log p(y) + F_M^0,
\end{equation}
where $p(y) = \sum_x p(y|x)p(x)$ and $F_M^0$ is the free
energy of $M$ in the standard state.
If $S$ is in state $x$ when the measurement is carried on,
then the average work
performed on $M$ during the contact is given by:
\begin{equation}
W_{\rm cont}(x) = kT\sum_{y} p(y)\log\frac{p(y)}{p(y|x)} +
\Delta F_M(x),
\end{equation}
where $\Delta F_M(x) = F_M^0 - F_M(x)$ is the change in free
energy of $M$ during the process. Hence, on average over
$x$, the work performed is
\begin{equation}
W_{\rm cont} = \sum_x p(x)W_{\rm cont}(x).
\end{equation}
The work performed during the separation, conditioned on
$x$ is given by
\begin{equation}
W_{\rm sep}(x) = kT \sum_y p(y|x)\log\frac{p(y|x)}{p(y)}
-\Delta F_M(x).
\end{equation}
As in the previous section, the average over $x$ of this
quantity is linked to the
mutual information between $x$ and $y$:
\begin{equation}
W_{\rm sep} = \sum_x p(x) W_{\rm sep}(x) = kT I - \Delta
F_M,
\end{equation}
where $\Delta F_M = \sum_x p(x) \Delta F_M(x)$ is the
average change in free energy of $M$ and $I$ is the mutual
information given by eq.~(\ref{eq:info}).
The total work performed is the sum of the two
contributions. It can be rewritten in following form:
\begin{equation}\label{eq:worktotgen}
W_{\rm tot} 
= kT \sum_{x,y}
\left( p(x,y) - p(x)p(y)\right)\log p(y|x).
\end{equation}

Let us now consider the second protocol. The measurement
device is left untouched after the previous measurement and
is thus in equilibrium with its energies set to $E_M(y|x')$
for a certain $x'$ appearing with probability $p(x')$. Here,
$x'$ is the state the system $S$ was during the previous
measurement. After the contact, the energies of $M$ are set
to $E_M(y|x)$, where $x$ is the current state of $S$,
appearing with probability $p(x)$ (independently of $x'$).
Given $x'$ and $x$, the average work performed on $M$ is:
\begin{equation}
W'(x',x) = kT\sum_y p(y|x')\log\frac{p(y|x')}{p(y|x)} +
F_M(x) - F_M(x').
\end{equation}
On average over $x'$ and $x$, we obtain following
expression:
\begin{eqnarray}
W' & = & \sum_{x',x}p(x')p(x) W(x',x) \nonumber\\
& = & kT\sum_{x',x,y}
p(x')p(x)p(y|x')\log\frac{p(y|x')}{p(y|x)}.
\end{eqnarray}
Using the fact that $p(y) = \sum_{x'} p(y|x')p(x')$, one can
bring the above expression to the same form as eq.
(\ref{eq:worktotgen}):
\begin{equation}
W' = kT \sum_{x,y}
\left( p(x,y) - p(x)p(y)\right)\log p(y|x) = W_{\rm tot}.
\end{equation}
This extends the result obtained in the previous section for
a two states system.
It means that switching $M$ from a
random state to another, or switching it from the standard
state to a random state and then back to the standard state
involves the same quantity of work on average.

\section{Relation to Landauer's principle}

In its original formulation, Landauer's principle states
that erasing one bit of information is necessarily
accompanied by the dissipation of at least $kT\log 2$ of
work \cite{landauer}. Bennett used Landauer's principle to
propose a solution to Szilard's paradox \cite{bennett}.
Let us briefly sketch the argument behind Landauer's
principle. Consider a one particle gas in a closed container in
contact with a heat bath. The volume is divided in two by a
removable partition. If the particle is in the left half of
the container, it encodes one value, say ``0'' and if it is
in the right half of the container, it encodes the other
value, ``1'' in this case.
Erasing the information contained in the memory means
bringing the memory to a standard state, say the state
``0'', without knowing the value that is encoded. Such a
protocol need to bring both states ``0'' and ``1'' to ``0''.
An obvious way to proceed is to remove the partition and
then compress the gas in the left half of the container.
Such a process dissipates at least $kT \log 2$. The
dissipation occurs during the free expansion of the gas
right after the removal of the partition: the volume of the
gas is doubled, i.e. its entropy is increased by $k\log 2$ and
no heat is exchanged with the heat bath meaning that the
entropy of the latter stays constant during this process.
Compressing the gas in the left half of the container needs
$kT\log 2$ of work which is then transferred to the heat
bath in form of heat, thereby increasing its entropy by
$k\log2$. This result was extended to general memories, see
e.g. \cite{sagawa2009}.

If the bit was used to store the result of the measurement
of the state of a two levels system, as in section
\ref{sec:setup}, then the work needed to erase the bit
surely compensates the maximum work that can be extracted
with help of the measurement because $I(p)\leq \log 2$ with
equality if and only if $p=1$ (or $p=0$), i.e. when there are
no measurement errors. In \cite{bennett}, Bennett argues
that the measurement can be performed reversibly. By this,
he means that as long as we know in which state the memory
is (say in the state encoding ``0'', the standard state), we
can reversibly drive it to ``0'' or to ``1'' according to
the result of the measurement. We will call this process
``recording of the information'' rather than ``
measurement''. However, this argument implies that the
system $S$ does not evolve during the recording, so that the
information is still useful when the process of recording
is finished. This is not the case in general. Consider for
instance the (imperfect) measurement of the position of a
Brownian particle as presented in \cite{abreu2011}: once one
has measured the position of the particle, one should
immediately perform a process depending on the measured
position in order to convert all of the information into
heat. In fact, the particle does not stop moving after the
measurement. So if one takes an infinite time to
reversibly record the information before using it, then once
it will be recorded it will bring nothing anymore because
the system will have relaxed back to equilibrium.

In our framework, the information about the result of the
measurement is stored in a way similar to the situation
considered in Landauer's principle: the measurement device
is in a probabilistic superposition of macroscopic states, on
per value of $x$ and appearing with probability $p(x)$.
However, there are two important differences.
We do not only consider dissipation during the erasure
step but also during the measurement itself. As
argued above, reversible recording of the
information implies that the fluctuation about which the
information is recorded is ``frozen''. If instead the system
is still in contact with the heat bath and continues to
evolve, then the driving of the measurement device needs to
be fast. However,  the main reason for the instantaneous
changes in the energies of $M$ is that it should be directly
driven by $S$: either the energies have a value that
directly depends on $x$ or not. But we consider no in
between. In our opinion, this is what makes the difference
between ``measurement'' and ``recording''. We consider
``recording'' to be the following process: we know the
outcome of some measurement and we want to record it to some
memory device. This can be done infinitely slowly and hence
with an arbitrarily small dissipation. On the other hand, a
``measurement'' device is directly driven by the information
which is measured, i.e. we do not have a direct control on
it.
 What then makes the information utilizable is the
fact that $M$ relaxes instantly. If $M$ is not fully
relaxed, then the information obtained is less, but the work
performed is the same.

Finally, let us remark that Landauer's erasure step is
analogous to our ``separation'' step. During this step, the
measurement device is brought from a mixture of macro-states
to the standard state similarly to Landauer's bit, which is
brought from a statistical mixture of ``0'' and ``1'' to the
standard state ``0''. And in fact, the work dissipated
during this process is $kTI$, i.e. it exactly
compensates the work extracted from the heat bath using the
information $I$. One big difference however, is the presence
of measurement errors. As already mentioned, in the
situation considered here, unlike in
the classical situation usually involved in Landauer's principle,
the different $p(y|x)$ may overlap for different values of
$x$.
Sagawa and Ueda extended Landauer's principle to the
situation with measurement errors \cite{sagawa2009}.
But in their framework, the measurement errors consist of
an erroneous recording of the information and the different
macro-states of the memory encoding the different
measurement outcome are still perfectly distinguishable.

\section{Conclusion}

We have presented a  very simple model for a measurement
device and a protocol for the measurement of thermal
fluctuations. The basic considerations motivating this model
are the following: the measurement device should be a
physical system and should obey the laws of
thermodynamics and its state should depend on the value
which is measured. In particular, the measurement errors
should at least include thermal fluctuations. In that
respect, the model presented here is minimal: the
measurement errors are only due to thermal fluctuations.
We showed that under these assumptions, the measurement
process itself already leads to dissipation of work in
addition to the dissipation due to Landauer's erasure
principle.

We also showed that the work performed on the measurement device
is the same, whether the measurement device is intialized in
and eventually brought back to a standard state, or is
simply left as it is at the end of the measurement process.

\bibliographystyle{phjcp}
\bibliography{$HOME/biblio}{}

\begin{thebibliography}{10}

\bibitem{szilard29}
{\sc L.~Szilard},
\newblock {\em Zeitschrift für Physik} {\bf 53}, 840 (1929).

\bibitem{toyabe2010}
{\sc S.~Toyabe}, {\sc T.~Sagawa}, {\sc M.~Ueda}, {\sc E.~Muneyuki}, and {\sc
  M.~Sano},
\newblock {\em Nature Physics} {\bf 6}, 988 (2010),
\newblock arXiv:1009.5287.

\bibitem{abreu2011}
{\sc D.~Abreu} and {\sc U.~Seifert},
\newblock {\em Euro. Phys. Lett.} {\bf 94}, 10001 (2011),
\newblock arXiv:1102.3826.

\bibitem{hasa1}
{\sc H.~H. Hasegawa}, {\sc J.~Ishikawa}, {\sc K.~Takara}, and {\sc D.~J.
  Driebe},
\newblock {\em Phys.~Lett.~A} {\bf 374}, 1001 (2010),
\newblock arXiv:0907.1569.

\bibitem{hasa2}
{\sc K.~Takara}, {\sc H.~H. Hasegawa}, and {\sc D.~J. Driebe},
\newblock {\em Phys.~Lett.~A} {\bf 375}, 88 (2010).

\bibitem{esp2011}
{\sc M.~Esposito} and {\sc C.~{Van den Broeck}},
\newblock {\em Eur. Phys. Lett.} {\bf 95}, 40004 (2011),
\newblock arXiv:1104.5165.

\bibitem{sagawa2008}
{\sc T.~Sagawa} and {\sc M.~Ueda},
\newblock {\em Phys. Rev. Lett.} {\bf 100}, 080403 (2008),
\newblock arXiv:0710.0956.

\bibitem{sagawa2010}
{\sc T.~Sagawa} and {\sc M.~Ueda},
\newblock {\em Phys. Rev. Lett.} {\bf 104}, 090602 (2010),
\newblock arXiv:0907.4914.

\bibitem{sagawa2011}
{\sc T.~Sagawa} and {\sc M.~Ueda},
\newblock Nonequilibrium thermodynamics of feedback control,
\newblock arXiv:1105:3262, 2011.

\bibitem{horo2010}
{\sc J.~M. Horowitz} and {\sc S.~Vaikuntanathan},
\newblock {\em Phys. Rev. E} {\bf 82}, 061120 (2010),
\newblock arXiv:1011.4273.

\bibitem{pon2010}
{\sc M.~Ponmurugan},
\newblock {\em Phys. Rev. E} {\bf 82}, 031129 (2010),
\newblock arXiv:1004.4311.

\bibitem{landauer}
{\sc R.~Landauer},
\newblock {\em IBM J.~Res.~Dev.} {\bf 5}, 183 (1961).

\bibitem{bennett}
{\sc C.~H. Bennett},
\newblock {\em Int.~J.~Theor.~Phys.} {\bf 21}, 905 (1982).

\bibitem{fahn}
{\sc P.~N. Fahn},
\newblock {\em Found.~of Phys.} {\bf 26}, 71 (1996).

\bibitem{sagawa2009}
{\sc T.~Sagawa} and {\sc M.~Ueda},
\newblock {\em Phys. Rev. Lett.} {\bf 102}, 250602 (2009),
\newblock arXiv:0809.4098.

\bibitem{cover}
{\sc T.~M. Cover} and {\sc J.~A. Thomas},
\newblock {\em Elements of Information Theory},
\newblock Wiley-Intersciences, second edition, 2006.

\bibitem{Ladyman2008}
{\sc J.~Ladyman}, {\sc S.~Presnell}, and {\sc A.~J. Short},
\newblock {\em Studies in History and Philosophy of Modern Physics} {\bf 39},
  315 (2008).

\bibitem{horo2011}
{\sc J.~M. Horowitz} and {\sc J.~M.~R. Parrondo},
\newblock {\em Eur. Phys. Lett.} {\bf 95}, 10005 (2011),
\newblock arXiv:1104.0332.

\bibitem{kawai2007}
{\sc R.~Kawai}, {\sc J.~M.~R. Parrondo}, and {\sc C.~{Van den Broeck}},
\newblock {\em Phys.~Rev.~Lett.} {\bf 98}, 080602 (2007),
\newblock arXiv:0707.2996.

\bibitem{gomez2008}
{\sc A.~Gomez-Marin}, {\sc J.~M.~R. Parrondo}, and {\sc C.~{Van den Broeck}},
\newblock {\em Euro.~Phys.~Lett.} {\bf 82}, 50002 (2008),
\newblock arXiv:0807.1027.

\end{thebibliography}
\end{document}